\documentclass[11pt,a4paper]{article}
\usepackage[T1]{fontenc}
\usepackage[utf8]{inputenc}
\usepackage[british]{babel}
\usepackage{amsmath,amssymb,amsfonts,bm}

\usepackage{enumitem,array}
\usepackage[margin=2.4cm]{geometry}
\usepackage{graphicx,float}
\usepackage{xcolor}
\definecolor{linkcol}{RGB}{30,70,120}
\usepackage[colorlinks=true,allcolors=linkcol]{hyperref}
\newcommand{\K}{\mathbf{K}}
\newcommand{\R}{\mathbf{R}}

\newcommand{\Mc}{\mathcal{M}}
\newcommand{\Lc}{\mathcal{L}}
\newcommand{\avg}[1]{\left\langle #1 \right\rangle}

\title{\textbf{The memory equation of a characteristic function in surface diffusion}}
\author{S. Miret-Art\'es\\[4pt]
\small Instituto de F\'isica Fundamental, CSIC, Serrano 123, 28006 Madrid, Spain}
\date{\today}

\begin{document}
\maketitle

\begin{abstract}
\noindent
The decay of a density wave, the Fourier component of the adparticle density seen
by a coherent probe, obeys in an interacting adlayer an exact memory equation of
Mori's form, determined by the local rate and the memory function. Because the
density is conserved, the correlation function is a characteristic function, and
the kernel is built from that of a single jump; positivity and normalization then
make the local rate and the memory function equilibrium averages, whatever the
interaction. The local rate obeys an exact sum rule, which makes de Gennes
narrowing a theorem on the lattice and gathers the interaction into one effective
hop rate. Detailed balance makes the decay a superposition of relaxation modes,
approached from below by a convergent hierarchy of closed forms. The theory fixes
the local rate, the memory function and the relaxation spectrum. The experiment provides the static structure factor and the decay, recorded from the instant of
scattering through a ballistic regime, a jump regime in which the memory acts,
and a diffusive regime. A single exponential fitted, as experiments do, over a time interval in the
diffusive regime gives the slowest relaxation rate, the constant rate of that
regime, and not the local rate. The correlation time, defined as the area beneath
the normalized decay, needs no time interval and is the mean inverse relaxation
rate. Under a
standard condition on the jump rates, met in one-dimensional diffusion when the rate of leaving a
site is independent of the site reached, the correlation time, the static structure factor and the lattice geometry give the effective hop rate at long wavelength,
with no model or fit and with a parameter-free test. There the sum rule becomes
a Darken-type relation.
\end{abstract}

\vspace{2pt}

\section{Introduction}
\label{sec:intro}

An adparticle, a particle adsorbed on a crystal surface, is bound to a site and
escapes from it intermittently; the solid supplies both the binding and the
thermal agitation which overcomes it. In a dilute adlayer the departures are
independent, and the coherent signal of a scattering experiment decays
exponentially at a rate fixed by the lattice and the temperature. Chudley and
Elliott established this for a liquid, and it holds on a periodic
substrate~\cite{Chudley1961,Montroll1965}. At finite coverage it fails. An
occupied site cannot receive an adparticle, the energy of a jump depends on the
occupied neighbouring sites, and the rate of departure depends on the
instantaneous configuration of the surroundings. The signal is no longer a single
exponential, except at long times, and the fitted rate is no longer the hop rate.

A coherent scattering experiment registers a density wave: the Fourier component
of the fluctuation of the adparticle density at one wavevector, the momentum
transfer parallel to the surface fixed by the geometry of the experiment. The
occupations which govern the departures do not appear in the density wave.
Eliminating them leaves a memory equation, of the kind Nakajima and Zwanzig
obtained by projection and Mori wrote for a single dynamical
variable~\cite{nakajima,zwanzig60,Mori1965,Zwanzig2001}. The equation is exact
and is determined by the local rate, the rate at which the density wave begins to
decay, and by the memory function, the delayed influence of the eliminated
occupations, which moderates the decay at later times. Both are commonly supplied
by a model, and every conclusion is then a property of the model.

This work has three goals. The first is to determine the local rate and the
memory function with no model of the adlayer. Four hypotheses are imposed:
conservation of the adparticles, a fixed set of jump vectors, detailed balance
and translational invariance. Every lattice gas which satisfies them belongs to
the class, and the results apply to each member, whatever the form and sign of
its lateral interaction; the range is also left open, except for one convergence
result, which requires it to be finite. The coverage is arbitrary: the results
hold from the dilute limit, where they reduce to the decay of Chudley and Elliott,
to coverages close to full occupation. The second goal is to establish the
content of a scattering measurement. A helium spin-echo
experiment~\cite{Jardine2009} fits a single exponential to the measured decay
over a window, a chosen range of times, and reports its amplitude and its rate,
the dephasing rate. From the instant of scattering, the measured signal passes
through a ballistic regime, a jump regime in which the memory of earlier jumps
acts, and a diffusive regime of dynamical equilibrium. The window always lies in
the diffusive regime, and the fit reports the constant rate of that regime. The
fitted amplitude and rate are not the local rate and the memory function, and the
discrepancy between them can be computed from the measured quantities. The third
goal is a procedure an experiment can carry out, with a test the data must pass,
using neither a model of the adlayer nor a fitted rate.

The novelty lies not in the equation but in the statements it allows about the
local rate and the memory function, and the reason is the projection variable. In
the construction of Mori and Zwanzig the variable is arbitrary, its correlation
function need only be square integrable, and the local rate and the memory
function are two functions about which no more can be said. Here the variable is
a Fourier component of a conserved density, so its correlation function is a
characteristic function (CF): of the separation in the momentum transfer and, as
will appear, of a spectrum of relaxation rates in the time. The kernel of the
equation is built from a second CF, that of a single jump. Positivity and
normalization are thereby available where the general formalism has only square
integrability, and detailed balance makes the generator self-adjoint and the
projections equilibrium averages, computable with no trajectory traced.

Four results are obtained. First, an exact sum rule relates the local rate to the
statics of the adlayer at every wavevector and for every member of the class, and
makes the narrowing described by de Gennes~\cite{deGennes1959} a theorem on the
lattice. Relations of the same algebraic form are known from mean-field
treatments~\cite{SinhaRoss1988,LeitnerVogl2011}; the sum rule makes them exact for
any interaction. Second, the decay is a superposition of relaxation modes whose
moments are static averages of the adlayer, and truncating the associated
continued fraction yields a convergent hierarchy of closed forms, each below the
exact decay at every time. Third, the fitted rate, the dephasing rate a spin-echo
experiment reports, is identified: for a window in the diffusive regime, where
experiments place it, it is the rate of the slowest relaxation mode, below the
local rate, which no fit reaches. Fourth, the area beneath the decay, with the
structure factor and the single-jump geometry, gives the effective hop rate at
long wavelength, with no adjustable parameter and with a test the data must pass.
This holds whenever the dynamics is of gradient type, a class standard in the
theory of lattice gases~\cite{Spohn1991,KipnisLandim1999} and the lattice
counterpart of Fick's law. In Fick's law the flow of particles between two points is set by the difference of their concentrations. Here the mean flow of adparticles between two neighbouring sites is the difference of one local quantity taken at the two sites; this quantity is built from the occupations around a site and is the same at every site. The simplest example is one-dimensional diffusion in which the rate of leaving a site depends on the occupations around it but not on the site reached, the jump being made only if that site is empty. On a two-dimensional lattice the same rule does not in general meet the condition. No
numerical results are given for any particular adlayer: the effective hop rate,
which gathers the interaction, is defined but not evaluated.

Three quantities enter these results, each for a definite reason. The correlation time $\tau_c$ is the area beneath the normalized decay. It is needed because a fit in the diffusive regime gives only the slowest relaxation rate, whereas the area takes in the whole decay, with no window and no assumed form. Its reciprocal lies between the slowest rate and the local rate. The width $\Delta$, the relative
spread of the relaxation rates, is needed because it fixes the initial value of
the memory function and the error of the simplest closed form. The ratio $R$, never below unity, is the factor by which the memory of the lateral interaction lengthens the correlation time with respect to a decay without memory. It is needed because the memory function cannot be measured, whereas $R$ is obtained from the measured decay and the static structure factor. Its value fixes how many orders of the hierarchy the data require.

Section~\ref{sec:cf} states the four hypotheses and shows that the intermediate
scattering function is a characteristic function, at dilute and at finite
coverage. Section~\ref{sec:eom} derives the memory equation, Sec.~\ref{sec:coeff}
the sum rule, and Sec.~\ref{sec:stieltjes} the relaxation spectrum and the
hierarchy of closed forms. Section~\ref{sec:measured} relates them to the
measured quantities and closes with the procedure. Section~\ref{sec:concl} sets
the theory and the experiment on a common time axis and states the matters left
open. An appendix gives the first orders explicitly.

\section{The intermediate scattering function as a characteristic function}
\label{sec:cf}

A coherent probe does not resolve individual adparticles. It registers the
Fourier component of the density wave labelled by $\K$, the momentum transfer
parallel to the surface, fixed by the geometry of the experiment, and records how
fast this component loses its coherence. The first Brillouin zone is called the
zone below, and diffusion across the zone means the behaviour at every $\K$ within
it, from the long-wavelength limit to the zone boundary.

Four hypotheses are used in the derivations:
\begin{enumerate}[label=(H\arabic*),leftmargin=3.2em]
\item\label{p:conserv}
\emph{Conservation.} The state of the adlayer is given by occupation numbers
$n_{\R}\in\{0,1\}$ (empty and occupied), one per point $\R$ of a Bravais lattice,
with mean $\avg{n_{\R}}=\theta$, the surface coverage; the dynamics moves
adparticles without creating or destroying them, so $\sum_{\R}n_{\R}$ is a
constant of the motion.
\item\label{p:jumps}
\emph{Jumps.} The elementary jump is an exchange between a site and one of its
neighbours, along a jump vector $\mathbf L$ drawn from a set fixed by the lattice;
longer jumps are admitted if their vectors belong to the same set.
\item\label{p:rev}
\emph{Reversibility.} The transition rates satisfy detailed balance with respect
to an equilibrium distribution over the configurations of the adlayer: the flux
from one configuration (a set of occupation numbers) to another equals the reverse
flux for every pair.
\item\label{p:inv}
\emph{Translational invariance.} The adlayer is invariant under the translations
of the lattice and under inversion; in particular the set of jump vectors and the
weights upon it are unchanged by $\mathbf L\to-\mathbf L$.
\end{enumerate}

The lateral interaction is left open: its form, range and sign, and the dependence
of the rates on the surrounding occupations beyond detailed balance. Only one
result, a statement of convergence, requires a finite range. The dimension of the
lattice is arbitrary. Every process satisfying \ref{p:conserv}--\ref{p:inv}
belongs to \emph{the class}, and every statement made of the class holds for all
of them, whatever the interaction.

The substrate acts as a thermal
bath~\cite{TorresMiyares2026a,TorresMiyares2026b,TorresMiyares2026c}. Adlayer and
substrate together are described by a density operator $\hat\rho_{\rm tot}(t)$,
and the trace over the substrate leaves the reduced density operator of the
adlayer,
\begin{equation}
	\hat\rho_{\rm red}(t) \;=\; {\rm Tr}_{\rm bath}\,\hat\rho_{\rm tot}(t) .
	\label{eq:rhored}
\end{equation}
Its evolution under the Liouville--von Neumann equation of the whole system is
exact at every coverage and for every lateral interaction, and only its diagonal
in the lattice basis $\{\,|\R\rangle\,\}$ enters the signal of a coherent probe.
At dilute coverage this diagonal distributes the displacement of one adparticle;
at finite coverage, the separation between two. The two cases are taken in turn.

\subsection*{Dilute coverage}

At dilute coverage the adparticles stand far apart, no site a jump requires is
occupied, and the departures are independent. The adlayer behaves as independent
copies of one adparticle, and the signal is built from its motion. Let
$|\R\rangle$ be the state in which it occupies the site $\R$, with the origin
occupied at $t=0$. The diagonal element
\begin{equation}
	\rho^{(1)}_{\R}(t) \;=\; \avg{\,\R\,|\,\hat\rho_{\rm red}(t)\,|\,\R\,}
	\label{eq:rho1}
\end{equation}
is the probability of finding it at $\R$ at time $t$, of unit weight
$\sum_{\R}\rho^{(1)}_{\R}(t)=1$ and starting from
$\rho^{(1)}_{\R}(0)=\delta_{\R,\mathbf 0}$.

The lattice point reached is the accumulated displacement, a sum of elementary
jumps,
\begin{equation}
	\R(t) \;=\; \sum_j n_j(t)\,\mathbf L_j ,
	\label{eq:jumps}
\end{equation}
with $\mathbf L_j$ the jump vector along the crystallographic direction $j$, of
length $a_j=|\mathbf L_j|$, and $n_j(t)$ the net number of jumps along it, forward
minus backward; the sum runs over every jump direction of the lattice, nearest
neighbours and longer jumps alike. Only the component of $\R$ along $\K$ enters
the scattered phase,
\begin{equation}
	\K\cdot\R(t) \;=\; \sum_j n_j(t)\,\K\cdot\mathbf L_j \;=\; K\,L_{\parallel}(t),
	\qquad
	L_{\parallel}(t) \;=\; \sum_j n_j(t)\,a_j\cos\tilde\beta_j ,
	\label{eq:Lpar}
\end{equation}
with $\tilde\beta_j$ the angle between $\mathbf L_j$ and $\K$. A direction
orthogonal to $\K$ drops out of \eqref{eq:Lpar} whatever the number of jumps along
it; more generally a direction is silent whenever $e^{\,i\K\cdot\mathbf L_j}=1$,
which a direction not orthogonal to $\K$ may satisfy at the zone boundary. Thus
$L_{\parallel}$ counts the directions the momentum transfer resolves, each weighted
by the cosine of its angle with $\K$; it is the variable the experiment resolves
at dilute coverage. The transform of \eqref{eq:rho1} is
\begin{equation}
	I^{(1)}(\K,t) \;=\; \sum_{\R} e^{\,i\K\cdot\R}\,\rho^{(1)}_{\R}(t)
	\;=\; \avg{\,e^{\,iKL_{\parallel}(t)}\,} ,
	\label{eq:cf1}
\end{equation}
the CF of the accumulated displacement in the strict sense, equal to unity at
$t=0$ and at $\K=\mathbf 0$. Its derivatives with respect to $K$ at fixed
direction generate the moments of $L_{\parallel}$, and those of its logarithm the
cumulants,
\begin{equation}
	\avg{L_{\parallel}^{\,n}(t)} \;=\; (-i)^n\,
	\frac{\partial^{\,n}I^{(1)}(\K,t)}{\partial K^{\,n}}\bigg|_{K=0},
	\qquad
	\kappa^{(1)}_n(t) \;=\; (-i)^n\,
	\frac{\partial^{\,n}\ln I^{(1)}(\K,t)}{\partial K^{\,n}}\bigg|_{K=0} .
	\label{eq:cumulants1}
\end{equation}
Forward and backward jumps being equally likely \ref{p:inv}, every odd order
vanishes and the second cumulant equals the second moment, a coincidence limited
to that order. It is a mean-square displacement, non-negative and zero at $t=0$,
which grows linearly at a rate fixing the diffusion coefficient of one adparticle,
and at dilute coverage of the adlayer. With independent jumps the decay is a
single exponential of unit amplitude, with no structure factor and no lateral
interaction in it.

\subsection*{Finite coverage}

At finite coverage the adparticle found at a lattice point is generally not the
one which started at the origin, and the count \eqref{eq:jumps} is no longer
available: it belongs to one adparticle, while the exchanges of \ref{p:jumps} move
them all. The diagonal of \eqref{eq:rhored} now distributes the separation
between an adparticle at the origin at time zero and an adparticle at time $t$. In
the occupation numbers of \ref{p:conserv},
\begin{equation}
	\rho_{\R}(t) \;=\;
	\frac{\bigl\langle\, n_{\mathbf 0}(0)\,n_{\R}(t)\,\bigr\rangle}{\theta} ,
	\label{eq:rhoR}
\end{equation}
the denominator being the probability that the origin is occupied. In a uniform
adlayer the origin is immaterial, and the separation alone labels the state.

Three properties of \eqref{eq:rhoR} are exact. First, $\rho_{\R}(t)\ge0$, as an
average of a product of occupation numbers. Second, at $t=0$ it is the equilibrium
separation distribution, $\rho_{\R}(0)=P_{\R}$, with $\rho_{\mathbf 0}(0)=1$ and
$\rho_{\R}(0)\to\theta$ at large separation, where sites are uncorrelated. Third,
its total weight is conserved \ref{p:conserv},
\begin{equation}
	\sum_{\R}\rho_{\R}(t) \;=\; \sum_{\R}\rho_{\R}(0) \;=\; N\theta
	\qquad\text{at all } t ,
	\label{eq:norm}
\end{equation}
with $N$ the number of sites of the lattice and $N\theta$ the number of
adparticles. Divided by this constant, $\rho_{\R}(t)$ is a probability
distribution at every time. All three properties differ from those of
\eqref{eq:rho1}: the weight is $N\theta$ and not unity, and the distribution at
$t=0$ is the equilibrium structure of the adlayer and not a point at the origin.

The coherent intermediate scattering function (ISF) is the lattice Fourier
transform of this distribution,
\begin{equation}
	I(\K,t) \;=\; \sum_{\R} e^{i\K\cdot\R}\,\rho_{\R}(t),
	\label{eq:cfdef}
\end{equation}
with $\K$ of modulus $K$ and fixed direction. Since \eqref{eq:rhoR} tends to
$\theta$ at large separation, \eqref{eq:cfdef} is a transform in the sense of
distributions, whose uniform background and decaying part give the two parts of
the signal. With the occupation fluctuation
\begin{equation}
	\delta n_{\R} \;\equiv\; n_{\R}-\theta ,
	\label{eq:deltan}
\end{equation}
the density wave of wavevector $\K$ is
\begin{equation}
	\delta n_{\K} \;=\; \sum_{\R}\delta n_{\R}\,e^{\,i\K\cdot\R} ,
	\label{eq:densitywave}
\end{equation}
and the two parts of the signal are
\begin{equation}
	I(\K,t) \;=\; \underbrace{\theta\sum_{\R} e^{\,i\K\cdot\R}}_{A(\K)}
	\;+\; \frac{\bigl\langle\,\delta n_{-\K}(0)\,\delta n_{\K}(t)\,
	\bigr\rangle}{N\theta} ,
	\label{eq:combsplit}
\end{equation}
the origin having been averaged over the $N$ equivalent sites. The first term is
a Dirac comb at the reciprocal-lattice points, the Bragg peaks: the elastic part,
fixed by the coverage, with no dynamics in it. The second converges absolutely
wherever the correlations of the adlayer are summable and contains the whole time
dependence: away from the Bragg points the experiment measures the equilibrium
autocorrelation of a single density wave. Its value at $t=0$ is the static
structure factor (SSF),
\begin{equation}
	I(\K,0) \;=\; F(\K) \;=\;
	\sum_{\R} \bigl[\,P_{\R}-\theta\,\bigr]\,e^{i\K\cdot\R},
	\label{eq:Fdef}
\end{equation}
itself the CF of the equilibrium separation distribution with its background
removed. The amplitude of \eqref{eq:cf1} is unity at every momentum transfer: the
SSF is the first mark the coverage leaves on the signal.

The jump vectors enter \eqref{eq:cfdef} through the phase a single exchange
displaces, not through a count of jumps. An exchange from $\R$ to $\R+\mathbf L$
turns $e^{\,i\K\cdot\R}$ into $e^{\,i\K\cdot(\R+\mathbf L)}$, and the density wave
changes by
\begin{equation}
	e^{\,i\K\cdot\R}\,\bigl(e^{\,i\K\cdot\mathbf L}-1\bigr) ,
	\label{eq:increment}
\end{equation}
which vanishes exactly when $e^{\,i\K\cdot\mathbf L}=1$. Only the other directions
act upon the signal, each through its projection $\K\cdot\mathbf L$. The geometry
of \eqref{eq:Lpar} survives while the trajectory which carried it does not, and
\eqref{eq:increment} is the quantity from which the local rate and the memory
function are built.

Normalized by its total weight $I(\mathbf 0,t)=N\theta$, \eqref{eq:cfdef} is a
CF in the strict sense at every time, of the separation and not of a
displacement. It generates the moments and the cumulants of the projection
$R_{\parallel}$ of the separation upon $\K$,
\begin{equation}
	\avg{R_{\parallel}^{\,n}(t)} \;=\; \frac{(-i)^n}{I(\mathbf 0,t)}
	\frac{\partial^{\,n}I(\K,t)}{\partial K^{\,n}}\bigg|_{K=0},
	\qquad
	\kappa_n(t) \;=\; (-i)^n\,
	\frac{\partial^{\,n}\ln I(\K,t)}{\partial K^{\,n}}\bigg|_{K=0} ,
	\label{eq:cumulants}
\end{equation}
which coincide with those of $L_{\parallel}$ only at dilute coverage. The odd
orders vanish by \ref{p:inv}, and the second cumulant again equals the second
moment, but no longer vanishes at the origin. Applied to the subtracted transform
of \eqref{eq:combsplit}, whose weight takes either sign, the second cumulant
separates into a static offset fixed by the structure of the adlayer and a part
growing linearly in the time,
\begin{equation}
	\kappa_2(t) \;=\; \ell_0^{2} \;+\; D_c\,t ,
	\qquad
	\ell_0^{2} \;=\; -\frac{\partial^{2}\ln F}{\partial K^{2}}\bigg|_{K=0} .
	\label{eq:kappa2}
\end{equation}
The offset may have either sign, being a cumulant of a signed weight, and the
slope $D_c$ is a collective transport coefficient, not the mobility of a labelled
adparticle. Transport is thus obtained from the ISF without constructing the
distribution of separations.

\subsection*{Time regimes}

The time $t$ is measured throughout from the instant of scattering, the origin a
spin-echo experiment uses, at which the signal has its static value $F(\K)$.
Several regimes succeed one another from this origin. For $t\ll\gamma^{-1}$, with
$\gamma$ the friction the substrate exerts on an adparticle, the adparticles move
freely: this is the ballistic regime, in which the signal leaves its initial value
as a Gaussian. It begins at the instant of scattering, $t=0$, with no interval
before it, and ends at $t\approx\gamma^{-1}$, when the adparticle has been deflected by the substrate. After the ballistic regime the motion is Brownian. In the corrugated potential of the substrate it
consists of a fast equilibration inside a well and of jumps between wells, and the
jump regime describes it on the scale of the jumps. The jump regime begins at the end of the ballistic regime and extends to long times; a trajectory is then a sequence of the jumps of
\ref{p:jumps} between wells, the motion inside a well being averaged over. At the
beginning of this regime the occupations around an adparticle have not yet
relaxed, and the lateral interaction delays the decay through a memory of earlier
jumps. As this memory decays, the jump regime reaches a dynamical equilibrium, the
diffusive regime, in which the decay proceeds at a constant rate. A description by jumps requires the ballistic regime and the equilibration inside a well to be short compared with the mean time an adparticle spends at a site, and this separation is assumed throughout. The
theory below describes the jump regime, its decay being continued down to $t=0$,
where it starts at $F(\K)$. The continuation is a property of the theory and not a
further regime: over the short interval $t\lesssim\gamma^{-1}$ it stands in for the
measured decay, ballistic and inside the well, which it does not describe.

Two of these regimes have a form known without any model of the adlayer. In the
ballistic regime, the expansion of \eqref{eq:cfdef} in the time has no first-order
term, forward and backward velocities being equally likely. At second order only the contribution of each adparticle with itself survives, since the velocities of distinct adparticles are uncorrelated in equilibrium. The signal therefore leaves the amplitude \eqref{eq:Fdef} quadratically in the time and in the momentum transfer~\cite{torres2026},
\begin{equation}
	I(\K,t) \;=\; F(\K)\;-\;\tfrac12\,\avg{v_{\parallel}^{2}}\,K^{2}t^{2}
	\;+\;O(t^{4}),
	\qquad t\ll\gamma^{-1} ,
	\label{eq:fsum}
\end{equation}
with $\avg{v_{\parallel}^{2}}$ the mean-square velocity along $\K$. For a
classical adparticle of mass $M$, equipartition gives
$\avg{v_{\parallel}^{2}}=k_BT/M$, with $T$ the surface temperature, the classical
frequency sum rule of a fluid~\cite{Hansen2013}; for a light adparticle the
velocity is the initial spreading velocity of the wave packet~\cite{torres2026}.
The friction enters neither case, the motion being free. The decay
\eqref{eq:fsum} does not depend on the structure of the adlayer and begins a
Gaussian decay,
\begin{equation}
	\frac{I(\K,t)}{F(\K)} \;\simeq\;
	\exp\left[-\,\frac{\avg{v_{\parallel}^{2}}\,K^{2}t^{2}}{2\,F(\K)}\right],
	\qquad t\ll\gamma^{-1},
	\label{eq:ballistic}
\end{equation}
Gaussian in the time, and also in the momentum transfer wherever the amplitude is
flat, as at dilute coverage, where the SSF is unity.

Equation \eqref{eq:fsum} is the classical f-sum rule. Let $S(\K,\omega)$ be the
Fourier transform in the time of the decay, extended evenly about the origin, so
$\int S\,d\omega=I(\K,0)$ and
$\int\omega^{2}S\,d\omega=-\partial^{2}I/\partial t^{2}|_{t=0}$, finite because
\eqref{eq:fsum} has zero slope at the origin. Then \eqref{eq:fsum} fixes the
second frequency moment of the measured spectrum,
\begin{equation}
	\frac{\int\omega^{2}\,S(\K,\omega)\,d\omega}{\int S(\K,\omega)\,d\omega}
	\;=\; \frac{\avg{v_{\parallel}^{2}}\,K^{2}}{F(\K)} ,
	\label{eq:fsumrule}
\end{equation}
fixed by the mass and the temperature and narrowed by the structure factor, with
no other trace of the lateral interaction. This is the moment de
Gennes~\cite{deGennes1959} considered, and the division by $F(\K)$ is the
narrowing he described. The sum rule derived below narrows the first moment of
the relaxation spectrum by the same factor, so both ends of the signal are
narrowed alike.

In the jump regime the adparticles have been deflected many times, and the
separation accumulates contributions the substrate has decorrelated. In the
diffusive regime the second cumulant of \eqref{eq:cumulants} grows linearly with
the time and the decay becomes a single exponential,
\begin{equation}
	I(\K,t) \;\simeq\; B(\K)\, e^{-\alpha(\K)\,t} .
	\label{eq:diffusive}
\end{equation}
A helium spin-echo (HeSE) measurement is fitted to \eqref{eq:diffusive} together
with the elastic baseline $A(\K)$ of \eqref{eq:combsplit}~\cite{Jardine2009}:
$\alpha(\K)$ is the dephasing rate it reports and $B(\K)$ the amplitude, which is
not the $F(\K)$ from which \eqref{eq:fsum} starts. The fit is always made over a
time window in the diffusive regime, and $\alpha(\K)$ is the constant rate of that
regime. The lateral interaction enters both numbers: neither the fitted amplitude
nor the fitted rate is a property of the adlayer alone, and neither is the rate at
which an adparticle leaves a site. An equation of motion for \eqref{eq:cfdef}
identifies them, and the sections below construct it.

\section{The memory equation}
\label{sec:eom}

From the end of the ballistic regime on, the distribution $\rho_{\R}$ evolves by
jumps, in the jump regime and in the diffusive regime which closes it. Whether an
adparticle can move to a neighbouring site, and how fast, depends on which
surrounding sites are occupied, and these occupations are absent from
$\rho_{\R}$: the separation is one variable of the adlayer among many, the others
having been removed in \eqref{eq:rhoR}. Eliminating them from the equation of
motion of $\hat\rho_{\rm red}$ by the projection of Nakajima and
Zwanzig~\cite{nakajima,zwanzig60} leaves a master equation for $\rho_{\R}$ alone,
in which the transfer between two lattice points is not instantaneous,
\begin{equation}
	\frac{\partial \rho_{\R}(t)}{\partial t}
	\;=\; \sum_{\R'}\int_0^t\! dt'\;\Bigl[\,
	\mathcal K_{\R-\R'}(t-t')\,\rho_{\R'}(t')
	\;-\;\mathcal K_{\R'-\R}(t-t')\,\rho_{\R}(t')\,\Bigr] ,
	\label{eq:gme}
\end{equation}
with $\mathcal K_{\R-\R'}(\tau)$ the transfer kernel, the rate at which an adlayer
presenting the separation $\R'$ a time $\tau$ earlier presents $\R$ now.
Conservation \ref{p:conserv} makes gain and loss enter with the same kernel and
opposite signs, since the sum of \eqref{eq:gme} over $\R$ must vanish, as
\eqref{eq:norm} requires; uniformity \ref{p:inv} makes the kernel depend on the
change of separation $\mathbf L=\R-\R'$ alone.

Equation \eqref{eq:gme} is exact because the equation of motion of
$\hat\rho_{\rm red}$ is linear and the adlayer starts in equilibrium
\ref{p:rev}, which removes the inhomogeneous term carrying the initial
correlations between the separation and the eliminated variables. A transfer made
now depends on earlier transfers, which left the occupations around the
adparticle not yet relaxed; this retardation is the memory acting in the jump
regime. In the Markovian limit the kernel has no delay,
$\mathcal K_{\mathbf L}(\tau)=\Gamma_{\mathbf L}\,\delta(\tau)$, with constant rates
$\Gamma_{\mathbf L}=\Gamma_{-\mathbf L}$, and \eqref{eq:gme} reduces to the gain
and loss balance
\begin{equation}
	\frac{\partial \rho_{\R}(t)}{\partial t}
	\;=\; \sum_{\mathbf L}\Gamma_{\mathbf L}
	\bigl[\,\rho_{\R+\mathbf L}(t)-\rho_{\R}(t)\,\bigr] ,
	\label{eq:pauli}
\end{equation}
the Pauli master equation for the diagonal elements, valid for thermally activated
jumps and for tunnelling alike, the two differing only in the rates. The sum runs
over the jump vectors of \ref{p:jumps}, and $\Gamma=\sum_{\mathbf L}\Gamma_{\mathbf L}$
is the total rate at which one adparticle leaves its site. Equation
\eqref{eq:pauli} is exact at dilute coverage, where every neighbouring site is
empty and the rates are constants; the retardation in \eqref{eq:gme} is the
contribution of the lateral interaction. The Markovian limit has no memory at any
time. It is not the diffusive regime of Sec.~\ref{sec:cf}, in which the kernel has
decayed after acting through the jump regime and the constant rate has already
been lowered by that action.

A lattice Fourier transform takes \eqref{eq:gme} to momentum-transfer space:
multiplying by $e^{\,i\K\cdot\R}$ and summing over $\R$ removes the coupling
between separations. In the gain term the substitution $\R=\R'+\mathbf L$
factorizes the double sum into a phase depending on $\mathbf L$ and the transform
\eqref{eq:cfdef}; in the loss term the sum over $\R'$ gives the same kernel at
zero momentum transfer. One scalar Volterra equation remains at each momentum
transfer,
\begin{equation}
	\frac{\partial I(\K,t)}{\partial t}
	\;=\; -\int_0^t\! dt'\; \mathcal W(\K,t-t')\, I(\K,t') ,
	\qquad
	\mathcal W(\K,\tau) \;=\; \sum_{\mathbf L}\bigl[\,1-e^{\,i\K\cdot\mathbf L}\,\bigr]
	\,\mathcal K_{\mathbf L}(\tau) .
	\label{eq:volterra}
\end{equation}
The factor $1-e^{\,i\K\cdot\mathbf L}$, the difference between loss and gain
phases, vanishes at $\K=\mathbf 0$ term by term, so $\mathcal W(\mathbf 0,\tau)=0$
and the weight of the signal stays constant: the dynamics respects the
conservation law of \eqref{eq:norm}. At $t=0$ the integral is empty and
$I(\K,0)=F(\K)$: the amplitude of the signal is fixed by the equilibrium adlayer,
not by its evolution.

One further CF is needed to interpret \eqref{eq:volterra}. Let $w_{\mathbf L}$ be
the probability that a single jump has the jump vector $\mathbf L$ of
\ref{p:jumps}, with $w_{\mathbf L}\ge0$ and $\sum_{\mathbf L}w_{\mathbf L}=1$, and
let
\begin{equation}
	f_{\K} \;=\; \sum_{\mathbf L} w_{\mathbf L}\, e^{\,i\K\cdot\mathbf L}
	\label{eq:fK}
\end{equation}
be the CF of this single-jump distribution at the momentum transfer of the
measurement; \eqref{eq:fK} is to one jump what \eqref{eq:cf1} is to the
accumulated displacement. Forward and backward jumps are equally likely,
$w_{\mathbf L}=w_{-\mathbf L}$, by the inversion in \ref{p:inv} and, independently,
by reversibility \ref{p:rev}, which equates the equilibrium fluxes between a pair
of sites in its two directions. Hence $f_{\K}$ is real and even in $\K$, and so is
every rate built upon it. In the Markovian limit $w_{\mathbf L}=\Gamma_{\mathbf L}/\Gamma$,
the kernel of \eqref{eq:volterra} is $\mathcal W(\K,\tau)=\Gamma[1-f_{\K}]\delta(\tau)$,
and the solution is a single exponential,
\begin{equation}
	I(\K,t) \;=\; F(\K)\,e^{-\Gamma\,[1-f_{\K}]\,t} ,
	\label{eq:ce}
\end{equation}
the decay of Chudley and Elliott~\cite{Chudley1961} with the static amplitude in
front; at dilute coverage $F(\K)\to1$, the form in which they wrote it. As a CF it
is the compound Poisson form of independent increments whose number in a time $t$
is Poisson distributed with mean $\Gamma t$~\cite{Montroll1965}. For
nearest-neighbour jumps along one lattice direction,
$f_{\K}=\cos(K a\cos\tilde\beta)$; longer jumps add higher harmonics.

The two contributions to $\mathcal W$ are not on the same footing: the jumps are
instantaneous on the scale of the interval between them, as the separation of time
scales of Sec.~\ref{sec:cf} requires. Part of the transfer therefore occurs with no
delay, and the dependence on earlier transfers resides in the rest. At each
momentum
transfer this separation is
\begin{equation}
	\mathcal W(\K,\tau) \;=\; \Omega(\K)\,\delta(\tau)\;-\;\Mc(\K,\tau) ,
	\label{eq:split}
\end{equation}
which defines $\Omega(\K)$ as the weight of the instantaneous part and
$\Mc(\K,\tau)$ as the remainder, the sign chosen so a positive memory function
slows the decay. Equation \eqref{eq:volterra} alone determines neither. With
\eqref{eq:split} it becomes
\begin{equation}
	\frac{\partial I(\K,t)}{\partial t}
	\;=\; -\,\Omega(\K)\, I(\K,t) \;+\; \int_0^t\! dt'\; \Mc(\K,t-t')\, I(\K,t') ,
	\label{eq:eom}
\end{equation}
the form Mori gave to the equation of motion of a single dynamical
variable~\cite{Mori1965,Zwanzig2001}, his projection and the one used here giving
the same equation. $\Omega(\K)$ is the local part of the transfer, acting before
any retardation has accumulated; at the beginning of the jump regime it is the
whole of the dynamics, the convolution being empty. Differentiating at $t=0$ the
decay of the jump regime, continued to the origin as in Sec.~\ref{sec:cf}, gives
the local rate alone,
\begin{equation}
	\Omega(\K) \;=\; -\,\frac{\partial \ln I(\K,t)}{\partial t}
	\bigg|_{t=0} ,
	\label{eq:Omega}
\end{equation}
the rate at which the decay sets out; the kernel acts only later, and a positive
memory can only slow the decay. The local term initiates the relaxation, the
retarded term moderates it. On the measured signal, which leaves the origin with
zero slope, \eqref{eq:fsum}, the logarithmic slope $-\partial\ln I/\partial t$
rises from zero through the ballistic regime and approaches $\Omega(\K)$ just
after it.

Equation \eqref{eq:eom} is exact under \ref{p:conserv}--\ref{p:inv}, and one
equation describes the whole jump regime, the diffusive regime included. At the
beginning of the jump regime only the local term acts; as the time grows the
convolution accumulates the memory and lowers the rate of decay; in the diffusive
regime the memory function has decayed, and the rate it leaves is constant. The ballistic regime and the equilibration inside a well lie outside the equation. Equation \eqref{eq:eom} differs from
the Mori equation of an arbitrary variable in governing a CF, and its kernel is
built from a second CF, \eqref{eq:fK}.

\section{A sum rule}
\label{sec:coeff}

The local rate and the memory function left open by \eqref{eq:eom} are properties
of the adlayer and require its configurations, not the separation alone. The
adlayer evolves by exchanges between an occupied site and an empty neighbouring
one~\cite{Kawasaki1966}, which conserve the number of adparticles.

By \eqref{eq:combsplit}, away from the Bragg points the signal is the equilibrium
autocorrelation of the single density wave \eqref{eq:densitywave}. Its mean-square
value per site is the density--density structure factor,
\begin{equation}
	S(\K) \;=\; \frac{\avg{|\delta n_{\K}|^{2}}}{N} ,
	\label{eq:SKdef}
\end{equation}
and the value of the signal at $t=0$ is fixed by the statics,
\begin{equation}
	I(\K,0)\;=\;\frac{\avg{|\delta n_{\K}|^{2}}}{N\theta}\;=\;\frac{S(\K)}{\theta}
	\;=\;F(\K) ,
	\label{eq:FSrel}
\end{equation}
which is \eqref{eq:Fdef} written in the occupations: $F(\K)$ refers to one
adparticle, $S(\K)$ to one site. Let $\Lc$ be the generator of the process, so a
quantity advanced by a time $t$ is $e^{\Lc t}$ acting upon it; \eqref{eq:gme} is
the same evolution with every variable but the separation eliminated.
Reversibility \ref{p:rev} means that the adlayer in equilibrium is statistically
the same whether the evolution runs forwards or backwards. It is a property of
the substrate driving the exchanges, not an assumption about the lateral
interaction, and it makes $\Lc$ self-adjoint and non-positive in the equilibrium
measure~\cite{liggett}. Since $\Lc$ advances the adlayer, \eqref{eq:combsplit}
becomes
\begin{equation}
	I(\K,t) \;=\; \frac{\avg{\,\delta n_{-\K}\;e^{\Lc t}\,\delta n_{\K}\,}}{N\theta},
	\label{eq:regression}
\end{equation}
an average over the equilibrium measure alone. Since $\Lc$ generates the jumps,
\eqref{eq:regression} is the decay of the jump regime, the diffusive regime
included, continued to the origin as in Sec.~\ref{sec:cf}: at $t=0$ it gives
$F(\K)$, and its slope there is that of the continuation.

The local rate and the memory function are obtained from \eqref{eq:regression} in two
steps. The first expresses both as equilibrium averages, by projecting the
instantaneous change of the density wave upon the density wave itself. The second
evaluates in closed form the average which gives the local rate, and yields the
sum rule.

The first step rests on a geometrical fact: the observables of the adlayer form a
vector space on which the equilibrium measure defines a scalar
product;\footnote{For two observables $A$ and $B$, functions of the configuration
$c$ of the adlayer,
\[(A,B)\;\equiv\;\avg{A^{*}B}\;=\;\sum_{c}p_{\rm eq}(c)\,A^{*}(c)\,B(c),\]
with $p_{\rm eq}(c)$ the equilibrium probability of $c$; it is linear in $B$,
conjugated by exchanging $A$ and $B$, and $(A,A)\ge0$. The density wave has length
$[N\theta F(\K)]^{1/2}$, and $(\delta n_{\K},A)=\avg{\delta n_{-\K}\,A}$.} $\Lc$ is
self-adjoint in this product, and the density wave spans a single direction. The
instantaneous change $\Lc\,\delta n_{\K}$ is itself an observable, and splits
uniquely into a component along $\delta n_{\K}$ and an orthogonal remainder,
\begin{equation}
	\Lc\,\delta n_{\K} \;=\; -\,\Omega(\K)\,\delta n_{\K}\;+\;\phi_{\K},
	\qquad
	\avg{\delta n_{-\K}\,\phi_{\K}} = 0 ,
	\qquad
	\Omega(\K) \;=\; -\,\frac{\avg{\delta n_{-\K}\,\Lc\,\delta n_{\K}}}
	{\avg{|\delta n_{\K}|^{2}}} .
	\label{eq:decomp}
\end{equation}
The local rate is minus the projection of $\Lc\,\delta n_{\K}$ upon the direction
of the measurement, real because $\Lc$ is self-adjoint and non-negative because
$\Lc$ is non-positive; the remainder $\phi_{\K}$ is the part of the instantaneous
dynamics the variable cannot express. Differentiating \eqref{eq:regression} at
$t=0$ recovers \eqref{eq:Omega}, so the two definitions of $\Omega(\K)$ agree, and
\eqref{eq:decomp} is an equilibrium average. The decomposition identifies the
memory function of \eqref{eq:eom} as the autocorrelation of the orthogonal
remainder,
\begin{equation}
	\Mc(\K,\tau) \;=\; \frac{\avg{\,\phi_{-\K}\;e^{\Lc_{\perp}\tau}\,\phi_{\K}\,}}
	{\avg{|\delta n_{\K}|^{2}}} ,
	\label{eq:memdef}
\end{equation}
with $\Lc_{\perp}$ the generator restricted to the space orthogonal to the density
wave; the operator identity behind \eqref{eq:memdef} is due to Mori, and its
derivation is standard~\cite{Mori1965,Zwanzig2001}. This memory belongs to the
lateral interaction: it acts through the jump regime; in the diffusive regime it
has decayed, leaving the constant rate of that regime. The friction of the
substrate, which ends the ballistic regime, is not part of it. Two consequences,
both resting on the self-adjointness and non-positivity $\Lc_{\perp}$ inherits
from $\Lc$, are used throughout. The memory function is non-negative --- itself a
superposition of decaying exponentials with non-negative weights, equal at zero
delay to $\Mc(\K,0)=\avg{|\phi_{\K}|^{2}}/\avg{|\delta n_{\K}|^{2}}$ --- and can
only oppose the decay, as the sign in \eqref{eq:split} displays. And its moments
are equilibrium averages, computable with no dynamics traced.

In the second step the numerator of \eqref{eq:decomp} is evaluated in closed form;
the denominator is known from \eqref{eq:FSrel}. The governing quantity is the
exchange frequency between a site and its neighbour along a jump vector,
\begin{equation}
J_{\mathbf L} \;=\; \bigl\langle\, w_{\R\to\R+\mathbf L}\;
n_{\R}\,\bigl(1-n_{\R+\mathbf L}\bigr)\,\bigr\rangle ,
\qquad
\Gamma_{\rm eff} \;=\; \frac{1}{\theta}\sum_{\mathbf L} J_{\mathbf L} ,
\label{eq:exfreq}
\end{equation}
the sum running over the directed jump vectors. Here $w_{\R\to\R+\mathbf L}$ is
the rate of the exchange, and the occupation factors record that it occurs only
from an occupied site to an empty one. Thus $J_{\mathbf L}$ is the mean number of
exchanges per unit time along $\mathbf L$ for a given pair of sites, all such
pairs being equivalent by \ref{p:inv}; the rate depends on the surrounding
occupations, and both factors are averaged over the equilibrium distribution.
Division by the coverage turns a frequency per site into a rate per adparticle:
$\Gamma_{\rm eff}$ is the mean rate at which one adparticle leaves its site.

The generator is a sum of terms, one for each directed jump. For a stationary
distribution, the real part of the numerator of \eqref{eq:decomp} reduces to a
sum over jumps, each contributing its rate times the squared change it produces in
the density wave, with no cross terms~\cite{liggett},
\begin{equation}
\mathrm{Re}\,\bigl\langle\,\delta n_{-\K}\,\Lc\,\delta n_{\K}\,\bigr\rangle
\;=\; -\sum_{\R}\sum_{\mathbf L}
\Bigl\langle\, w_{\R\to\R+\mathbf L}\, n_{\R}
\bigl(1-n_{\R+\mathbf L}\bigr) \Bigr\rangle\,
\bigl[\,1-\cos(\K\cdot\mathbf L)\,\bigr] ,
\label{eq:resplit}
\end{equation}
with the geometry already outside the average. A jump removes a unit of
occupation at the departure site and adds one at the target, changing the density
wave by $e^{\,i\K\cdot\R}\bigl(e^{\,i\K\cdot\mathbf L}-1\bigr)$. The phase of the
departure site cancels in the square, leaving $2\bigl[1-\cos(\K\cdot\mathbf L)\bigr]$:
a number, not a function of the state, which leaves the average. The factor two
cancels against the one half of a sum counting each pair of sites in both
directions.

Only the rate is left to be averaged, and its average is \eqref{eq:exfreq}.
Summing over the $N$ sites and using $\avg{|\delta n_{\K}|^{2}}=N\theta F(\K)$,
\begin{equation}
\mathrm{Re}\,\bigl[\Omega(\K)F(\K)\bigr]
\;=\; \Gamma_{\rm eff}\,\bigl[\,1-\mathrm{Re}\,f_{\K}\,\bigr] ,
\label{eq:dirichlet}
\end{equation}
with $f_{\K}$ the single-jump CF \eqref{eq:fK} built on the weights
\begin{equation}
w_{\mathbf L} \;=\; \frac{J_{\mathbf L}}{\sum_{\mathbf L'}J_{\mathbf L'}} .
\label{eq:weights}
\end{equation}
Equation \eqref{eq:weights} fixes these weights in general: \ref{p:jumps} names
the jump vectors, and \eqref{eq:weights} gives each its share of the exchanges, the
$w_{\mathbf L}=\Gamma_{\mathbf L}/\Gamma$ of Sec.~\ref{sec:eom} being the case of
constant rates.

Detailed balance is not used in reaching \eqref{eq:dirichlet}: stationarity alone
fixes the momentum dependence of the rate. By \ref{p:rev} the stationary
distribution is the equilibrium one, in whose scalar product $\Lc$ is
self-adjoint~\cite{liggett}.\footnote{With
$(\Lc B)(c)=\sum_{c'}w(c\to c')\,[B(c')-B(c)]$, $w(c\to c')$ being the transition
rate, detailed balance, $p_{\rm eq}(c)\,w(c\to c')=p_{\rm eq}(c')\,w(c'\to c)$,
gives
\[\avg{A^{*}\Lc B}\;=\;-\tfrac12\sum_{c,c'}p_{\rm eq}(c)\,w(c\to c')\,\bigl[A(c')-A(c)\bigr]^{*}\bigl[B(c')-B(c)\bigr],\]
symmetric in $A$ and $B$ up to conjugation (self-adjointness), never positive for
$A=B$, and equal to \eqref{eq:resplit} for $A=B=\delta n_{\K}$.} The expectation
value of a self-adjoint operator is real, so the imaginary part of the numerator
vanishes and \eqref{eq:dirichlet} is the whole product,
\begin{equation}
\Omega(\K)\,F(\K)
\;=\; \Gamma_{\rm eff}\,\bigl[\,1-f_{\K}\,\bigr] ,
\label{eq:sumrule}
\end{equation}
the sum rule. It holds for every process obeying \ref{p:conserv}--\ref{p:inv}: the
form of the interaction, its range, and any closed form for the structure factor
are never used.

Its two sides are of different kinds. On the left stands a property of the decay:
the rate at which the decay of the jump regime sets out, before the memory acts,
times the amplitude from which it starts. On the right stands a property of the
adlayer at rest: how often an adparticle reaches a neighbouring site, times a
factor fixed by the geometry of the jump. The product is fixed by the statics, and
its wavevector dependence is that of a single jump. The interaction does not act
on the product but on the two factors: it raises the amplitude wherever the
adlayer has organized itself and lowers the rate in the same proportion. Neither
$\Omega$ nor $F$ follows from the statics alone; their product does.

Rearranged for the rate, \eqref{eq:sumrule} separates it into a structureless
numerator and a structured denominator: relaxation slows wherever the adlayer has
built up structure and quickens where structure is suppressed. This is the
narrowing described by de Gennes~\cite{deGennes1959} for the coherent line width
of a dense fluid, here a theorem on the lattice, since \eqref{eq:sumrule} fixes
$\Omega(\K)F(\K)$ without using the form of $F(\K)$. Relations of the same
algebraic form were obtained for interstitials in a crystal by Sinha and
Ross~\cite{SinhaRoss1988}, in a self-consistent treatment of the random-phase
kind, and independently by Leitner and Vogl~\cite{LeitnerVogl2011}, in an
expansion about high temperature. Both close the numerator with the mean-field
blocking factor of a dilute occupancy, and the simulations of Leitner and Vogl show
the relation to fail where the interaction is strong. The sum rule adds three
things: the numerator is the exact exchange frequency \eqref{eq:exfreq}, the
interaction is unrestricted, and the quantity fixed is the first moment, as the
one de Gennes~\cite{deGennes1959} considered is the second, \eqref{eq:fsumrule}.
The two narrowings act at the two ends of the time axis, the f-sum rule in the
ballistic regime and the sum rule at the beginning of the jump regime, through the
same factor $1/F(\K)$. The interaction survives in $\Gamma_{\rm eff}$, which
\eqref{eq:sumrule} defines without evaluating.

The weights \eqref{eq:weights} are ratios of exchange frequencies, each depending
on the interaction through \eqref{eq:exfreq}. Where the symmetry of the adlayer makes all jump vectors equivalent, the weights are fixed by the lattice alone. Where inequivalent vectors coexist, a nearest neighbour and a longer jump, the interaction redistributes the weight between them, and $f_{\K}$ acquires a wavevector dependence of its own. The sum rule is untouched, but taking the
product $\Omega F$ to be a structureless rate times a lattice factor requires
equivalent jump vectors. The weights being even, $f_{\K}$ is real and
\begin{equation}
1-f_{\K} \;=\; \sum_{\mathbf L} w_{\mathbf L}\,
\bigl[\,1-\cos(\K\cdot\mathbf L)\,\bigr] ,
\label{eq:additive}
\end{equation}
each vector entering through its own projection $\K\cdot\mathbf L$, the sum
running over the jump vectors and not over a basis. The additivity holds for the
product $\Omega F$ and not for the decay, except at dilute coverage, where the
decay is the exponential of the product and factorizes into one function of each
projection. At finite coverage the memory \eqref{eq:memdef} is the autocorrelation of a residual $\phi_{\K}$ to which every jump contributes, since the occupations which block a jump in one direction also block the others. Neither the decay nor any rate drawn from it then separates by direction.

At long wavelength the adlayer can flatten a density wave only by transporting
adparticles from crests to troughs, and by \ref{p:conserv} the rate of flattening
vanishes with the wavevector. Expanding $f_{\K}$ for small momentum transfer, the
odd orders vanishing by \ref{p:inv}, with $F(0)$ the long-wavelength limit of the
structure factor, the sum rule gives
\begin{equation}
\Omega(\K) \;\xrightarrow[K\to0]{}\; D_c\,K^{2} ,
\qquad
D_c \;=\; \frac{\Gamma_{\rm eff}\,\avg{\ell^{2}}}{2\,F(0)} ,
\qquad
\avg{\ell^{2}} \;=\; \sum_{\mathbf L} w_{\mathbf L}\,
\bigl(\hat\K\cdot\mathbf L\bigr)^{2} ,
\label{eq:Dc1}
\end{equation}
with $\hat\K$ the unit vector along the momentum transfer and $\avg{\ell^{2}}$ the
mean squared projected jump length, a property of the geometry alone. Equation
\eqref{eq:Dc1} has the structure of the relation Darken obtained for
interdiffusion in alloys~\cite{Darken1948}, a mobility times a thermodynamic
factor. The factor is $1/F(0)$, the inverse compressibility by
\eqref{eq:regression}, and its appearance is the long-wavelength limit of the
narrowing. The numerator differs: Darken identifies it with the tracer diffusion
coefficient, while here it is $\Gamma_{\rm eff}\avg{\ell^{2}}/2$, the coefficient
of an adparticle hopping at the effective rate with no correlation between
successive jumps. The coefficient \eqref{eq:Dc1} belongs to the local rate, at the
beginning of the jump regime; the coefficient of the diffusive regime, the slope
in \eqref{eq:kappa2}, equals it only if the memory leaves the rate at long
wavelength unchanged.

\section{The relaxation spectrum}
\label{sec:stieltjes}

Equation \eqref{eq:eom} is a Volterra equation of convolution type, and its
Laplace transform, taken over the decay of the jump regime continued to the
origin, is exactly
\begin{equation}
\hat I(\K,s) \;=\; \frac{F(\K)}{s+\Omega(\K)-\hat\Mc(\K,s)} .
\label{eq:laplace}
\end{equation}
By \ref{p:rev}, the local rate and the memory function are generated by one
non-negative measure, which determines \eqref{eq:laplace}.

Reversibility fixes the form of this measure. The generator is self-adjoint and
non-positive in the scalar product of Sec.~\ref{sec:coeff}, so it has a complete
set of eigenvectors with real, non-negative relaxation rates. Expanded upon them,
the decay is a superposition of relaxation modes,
\begin{equation}
\frac{I(\K,t)}{F(\K)} \;=\; \sum_{k} a_k(\K)\,e^{-\lambda_k(\K)\,t} ,
\qquad a_k(\K) \;\ge\; 0 , \qquad \sum_k a_k(\K) \;=\; 1 ,
\label{eq:modes}
\end{equation}
with $\lambda_k$ the rate of the $k$th mode, an eigenvalue of $-\Lc$, and $a_k$ the
squared overlap of its eigenvector with the density wave: non-negative, and summing
to unity since the normalized decay starts at one. This is the decay of the jump
regime, the diffusive regime included, continued to the origin; the ballistic regime, in which the signal leaves its initial value as a Gaussian, and the equilibration inside a well lie outside it. In general no single mode accounts for the decay, and the measurement sees the
whole superposition, not the rates one by one. The weights define a measure on the
rates,
\begin{equation}
d\mu_{\K}(\lambda) \;=\; \sum_k a_k(\K)\,
\delta\bigl(\lambda-\lambda_k(\K)\bigr)\,d\lambda ,
\label{eq:mudef}
\end{equation}
a weight $a_k$ at each rate $\lambda_k$, $\delta$ being the Dirac delta. This is
the relaxation spectrum of the density wave. Written as a distribution, it also
covers the continuum of rates of a macroscopic adlayer; it is non-negative and of
unit weight, by \eqref{eq:modes}. Averages over it bear a subscript,
\begin{equation}
\avg{\,g(\lambda)\,}_{\mu} \;\equiv\; \int_0^{\infty}\! g(\lambda)\,
d\mu_{\K}(\lambda) \;=\; \sum_k a_k(\K)\,g\bigl(\lambda_k(\K)\bigr) ,
\qquad \avg{1}_{\mu} \;=\; 1 ,
\label{eq:muavg}
\end{equation}
unlike the averages over the adlayer. In this notation \eqref{eq:modes} and its
Laplace transform become
\begin{equation}
\frac{I(\K,t)}{F(\K)} \;=\; \avg{\,e^{-\lambda t}\,}_{\mu} ,
\qquad
\frac{\hat I(\K,s)}{F(\K)} \;=\; \Bigl\langle\,\frac{1}{s+\lambda}\,
\Bigr\rangle_{\mu} ,
\label{eq:spectral}
\end{equation}
the spectral representation. Expanding the exponential gives the moments of the
spectrum as static averages of the adlayer,
\begin{equation}
m_n(\K) \;\equiv\; \avg{\,\lambda^{n}\,}_{\mu}
\;=\; \frac{\avg{\,\delta n_{-\K}\,(-\Lc)^{n}\,\delta n_{\K}\,}}{N\theta\,F(\K)} ,
\label{eq:moments}
\end{equation}
so $m_0=1$ and, by \eqref{eq:decomp}, $m_1=\avg{\lambda}_{\mu}=\Omega(\K)$: the
local rate is the mean relaxation rate of the density wave. By
\eqref{eq:spectral}, the decay depends on the modes only through these moments,
and the modes need never be resolved individually.

The moments $m_n$ are not the derivatives at $t=0$ of the decay of the jump
regime, continued to the origin, which include the amplitude as well,
\begin{equation}
	(-1)^n\frac{\partial^{\,n}I(\K,t)}{\partial t^{\,n}}\bigg|_{t=0}
	\;=\; F(\K)\,m_n(\K) ,
	\label{eq:omegan}
\end{equation}
nor the frequency moments of the measured spectrum, the second of which is fixed
by the f-sum rule \eqref{eq:fsumrule}. That rule belongs to the ballistic regime,
where the decay leaves $F(\K)$ with zero slope; the decay \eqref{eq:modes} leaves
it with the slope $-F\Omega$, and its second frequency moment diverges, as for a
Lorentzian. The two sets constrain different regimes, and neither follows from the
other. The equation of motion fixes $m_1$ and $m_2$. At $t=0$ the convolution in
\eqref{eq:eom} is empty, giving $m_1=\Omega(\K)$ again; a second derivative at
$t=0$, where the convolution contributes $\Mc(\K,0)$, gives
\begin{equation}
	m_2(\K) \;=\; \Omega^{2}(\K)+\Mc(\K,0) ,
	\label{eq:om2}
\end{equation}
so the memory function starts at the spread of the rates about their mean,
\begin{equation}
	\Mc(\K,0) \;=\; \Delta(\K)\,\Omega^{2}(\K),
	\qquad
	\Delta(\K) \;=\; \frac{m_2}{m_1^{2}}-1
	\;=\; \frac{\avg{\lambda^{2}}_{\mu}-\avg{\lambda}_{\mu}^{2}}
	{\avg{\lambda}_{\mu}^{2}}
	\;\ge\; 0 .
	\label{eq:Delta}
\end{equation}
$\Delta(\K)$, the variance of the rates over the square of their mean, is their
squared relative width, non-negative because the weights are. It vanishes if and
only if one rate takes the whole signal. An adlayer relaxing through a single mode
has no memory and decays as one exponential; when several modes share the signal,
the mixture is not an exponential, and the memory function accounts for the
departure from one.

The Laplace transform of the decay is not an arbitrary function of $s$. Expanding
\eqref{eq:spectral} in powers of $1/s$ generates the moments of the spectrum,
\begin{equation}
\frac{\hat I(\K,s)}{F(\K)} \;\sim\;
\sum_{n\ge0}\frac{(-1)^{n}\,m_n(\K)}{s^{\,n+1}} ,
\qquad s\to\infty ,
\label{eq:mgf}
\end{equation}
so the transform is the moment-generating function of the relaxation spectrum,
the expansion being asymptotic in general and convergent beyond the largest rate
for a measure of bounded support~\cite{Widder1941}. A moment-generating function
with a non-negative measure can be written as a continued fraction with
non-negative coefficients~\cite{Baker1996,Akhiezer1965}, and for
\eqref{eq:laplace} it is the fraction Mori obtained by iterating the
projection~\cite{Mori1965,Lee1982},
\begin{equation}
\frac{\hat I(\K,s)}{F(\K)} \;=\;
\cfrac{1}{s+\Omega-\cfrac{b_1}{s+\Omega_1-\cfrac{b_2}{s+\Omega_2-\cdots}}} ,
\qquad b_n \;\ge\; 0 ,
\label{eq:cf}
\end{equation}
with coefficients $\Omega_n(\K)$ and $b_n(\K)$ generated by the moments
\eqref{eq:moments}. Thus $\Omega_0=\Omega$ is the local rate and $b_1=\Mc(\K,0)$ the initial value of the memory function. Each $\Omega_n$ is the mean rate of the $n$th residual and each $b_n$ its squared amplitude, non-negative as a property of the measure, not as an assumption on the dynamics. Truncating \eqref{eq:cf} at
order $n$, by setting $b_n=0$, leaves a ratio of polynomials in $s$ of degrees
$n-1$ and $n$: the Pad\'e approximant of \eqref{eq:mgf} reproducing the most
terms, the moments $m_0$ to $m_{2n-1}$~\cite{Baker1996}.

The denominator of this ratio is the characteristic polynomial of the $n\times n$
tridiagonal matrix $\mathsf J_n$ with $\Omega_0,\dots,\Omega_{n-1}$ on its
diagonal and $\sqrt{b_1},\dots,\sqrt{b_{n-1}}$ beside it; its $n$ roots are real,
positive and simple. Partial fractions inverted term by term give the decay at
order $n$,
\begin{equation}
I_n(\K,t) \;=\; F(\K)\sum_{j=1}^{n} a^{(n)}_j(\K)\,
e^{-\lambda^{(n)}_j(\K)\,t} ,
\qquad
a^{(n)}_j \;=\; \bigl|\,u_{j1}\,\bigr|^{2} ,
\qquad
\sum_{j=1}^{n} a^{(n)}_j \;=\; 1 ,
\label{eq:In}
\end{equation}
a sum of $n$ exponentials whose rates $\lambda^{(n)}_j$ are the eigenvalues of
$\mathsf J_n$, with $u_{j1}$ the first component of the corresponding normalized
eigenvector.

Equation \eqref{eq:In} has the form of \eqref{eq:modes} but is a different decay.
In \eqref{eq:modes} the sum runs over all the relaxation modes, with the true rates
and weights; in \eqref{eq:In} it has $n$ terms, the nodes and weights of a
quadrature fixed by the first $2n$ moments, not $n$ of the true rates and weights.
The two coincide only for a spectrum of at most $n$ rates; otherwise
\eqref{eq:In} approximates \eqref{eq:modes}. A measure of $n$ points sharing the
first $2n$ moments of $\mu_{\K}$ exists and is unique: its rates are the zeros of
the polynomial of degree $n$ orthogonal with respect to $\mu_{\K}$, and $I_n$ is
the decay it generates. This is the Gauss quadrature of the
spectrum~\cite{Gautschi2004}, and each order is computed by it. The hierarchy is
not a sequence of physical assumptions but the sequence of measures of finite
support sharing the low moments of the exact one, each computable from $2n$ static
averages of the adlayer.

The memory function is truncated in the same way. Comparing \eqref{eq:cf} with
\eqref{eq:laplace} identifies its transform as the remainder of the fraction,
\begin{equation}
\hat\Mc(\K,s) \;=\;
\cfrac{b_1}{s+\Omega_1-\cfrac{b_2}{s+\Omega_2-\cdots}} ,
\label{eq:memcf}
\end{equation}
so a truncation at order $n$ leaves for the memory a sum of $n-1$ exponentials.
The first order sets $\Mc=0$, discarding the memory. The second gives a single
exponential,
\begin{equation}
\Mc(\K,\tau) \;=\; b_1(\K)\,e^{-\Omega_1(\K)\,\tau} ,
\label{eq:memlevel2}
\end{equation}
the form customarily assumed when a memory function is modelled; here its
parameters are fixed by static averages, the initial value $b_1=\Mc(\K,0)$ exactly
and the decay rate $\Omega_1$ as the mean rate at which the dynamics orthogonal to
the density wave relaxes. At order $n$ the truncated \eqref{eq:memcf} is a ratio
of polynomials whose denominator has $n-1$ simple roots, and inverting it as for
the decay gives
\begin{equation}
\Mc_n(\K,\tau) \;=\; \sum_{j=1}^{n-1} c_j(\K)\,e^{-\nu_j(\K)\,\tau} ,
\qquad
\sum_{j} c_j \;=\; b_1 \;=\; \Mc(\K,0) ,
\label{eq:memtime}
\end{equation}
where the rates $\nu_j$ are the eigenvalues of the tridiagonal matrix
$\mathsf J_{n-1}$ with $\Omega_1,\dots,\Omega_{n-1}$ on its diagonal and
$\sqrt{b_2},\dots,\sqrt{b_{n-1}}$ beside it, and
\begin{equation}
c_j \;=\; b_1\,\bigl|\,v_{j1}\,\bigr|^{2} ,
\label{eq:memweights}
\end{equation}
with $v_{j1}$ the first component of the $j$th normalized eigenvector. The amplitudes are non-negative, so the truncated memory is non-negative at every delay and can only oppose the decay. They sum to $b_1$, so $\Mc(\K,0)$ is exact at every order beyond the first, and the approximation lies entirely in the shape of the subsequent relaxation. The second and third orders are written out in
Appendix~\ref{app:levels}. The hierarchy thus gives closed forms for the memory
as much as for the decay, and the customary exponential ansatz is its second
member, not an independent assumption.

Truncating at $n=1$ places the whole weight at the single rate
$\lambda=\Omega(\K)$, which the sum rule \eqref{eq:sumrule} fixes exactly:
\begin{equation}
I_1(\K,t) \;=\; F(\K)\,e^{-\Omega(\K)\,t} .
\label{eq:level1}
\end{equation}
This closed form is not an approximation adopted for convenience but the first
member of \eqref{eq:cf}, using the two moments the sum rule supplies. Its rate is
the single-jump factor $1-f_{\K}$ of Chudley and Elliott divided by the structure
factor, and this division separates the two forms. Theirs is, in the momentum
transfer, the CF of a compound Poisson displacement, as the constancy of the rate
in front of $1-f_{\K}$ requires. The first order is not, although the decay it
approximates is: after division by $F(\K)$ the exponent need not have an
everywhere non-negative inverse transform on the lattice, a failure of the
truncation and not of \eqref{eq:cfdef}. The interaction enters twice, raising the
amplitude and lowering the rate, and the two forms agree only where $F(\K)$ does
not depend on the momentum transfer, at dilute coverage.

Two properties of the first order are used below. It is a rigorous lower bound on
the decay at every momentum transfer and every time, and its departure from the
decay is fixed, as $t\to0$, by \eqref{eq:Delta} alone. Writing
$\lambda=\Omega(1+x)$ in \eqref{eq:spectral} and expanding in $x$, of zero mean
and mean square $\Delta$,
\begin{equation}
\frac{I(\K,t)-I_1(\K,t)}{F(\K)} \;=\;
\tfrac{1}{2}\,\Delta(\K)\,(\Omega t)^{2}\,e^{-\Omega t}
\;+\;O\bigl((\Omega t)^{3}\bigr) ,
\label{eq:level1error}
\end{equation}
a correction of first order in $\Delta$ and second order in $\Omega t$. The first
order has no memory: it agrees with the decay at the beginning of the jump regime
and falls below it as the memory acts, by \eqref{eq:level1error}.

Three rates must be kept apart, each with its place on the time axis of
Sec.~\ref{sec:cf}. The rates $\lambda_k$ belong to the individual modes of
\eqref{eq:modes}; the smallest with non-zero weight, $\lambda_{\min}(\K)$, governs
the decay at long times. The local rate $\Omega(\K)$ is their mean,
\eqref{eq:moments}, the single rate of the first order \eqref{eq:level1}, and the
rate at which the decay of the jump regime sets out, before the memory acts. The
rate $\alpha(\K)$ of \eqref{eq:diffusive} is the one a single exponential fitted to
the data reports, always over a window in the diffusive regime. The first order is
Markovian in the sense of Sec.~\ref{sec:eom}: it discards the memory at every time,
not only at late times. Its rate is $\Omega(\K)=\Gamma_{\rm eff}[1-f_{\K}]/F(\K)$,
which reduces to the rate $\Gamma[1-f_{\K}]$ of the Markovian limit \eqref{eq:ce}
at dilute coverage.

The diffusive regime of Sec.~\ref{sec:cf} can now be placed among these rates.
Each mode decays at its own rate, so the faster modes die out first, and the
logarithmic slope $-\partial\ln I/\partial t$ of the decay of the jump regime
falls monotonically from $\Omega(\K)$ at $t=0$ towards $\lambda_{\min}(\K)$; this
fall is the action of the memory. When $\lambda_{\min}$ is isolated from the other rates, as in a finite adlayer, the slope settles at $\lambda_{\min}$. This is the diffusive regime, in which the decay is the single exponential \eqref{eq:diffusive} with $\alpha=\lambda_{\min}$, the constant rate the memory leaves after acting through the jump regime. In a macroscopic adlayer the rates form a continuum, and the decay is not a pure exponential even at long times. Its logarithmic slope approaches the lower edge of the continuum only gradually, the diffusive regime is reached only asymptotically, and a fit there gives a rate close to that edge. At long wavelength one mode carries almost the whole weight, and the
distinction disappears. Where $\Delta(\K)$ vanishes, the three rates coincide: the
spectrum is a single rate, the decay is \eqref{eq:level1} exactly, at every time
and not merely to leading order in the interaction, and every fit returns
$\alpha=\Omega$. Where $\Delta(\K)$ does not vanish, the closed form remains
accurate over the decay provided $\Delta$ is small: at $\Omega(\K)t=1$, an excess
of ten per cent corresponds, to leading order, to $\Delta(\K)\simeq0.2$. The
accuracy of the closed form thus rests on one static number at each momentum
transfer, an equilibrium property of the adlayer.

Every truncation lies below the decay: a Gauss quadrature underestimates the
average of a function whose derivative of order $2n$ is positive, as that of
$e^{-\lambda t}$ is, and the truncations rise towards the decay as moments are
added. Convergence to the decay itself, and not to another decay sharing its
moments, requires bounded rates. They are bounded whenever they depend on the
occupations of a bounded neighbourhood, whose configurations are finite in number;
an interaction of finite range suffices, and is assumed from here on. By
\ref{p:jumps} an adparticle has finitely many directions in which to leave, and the
relaxation spectrum has compact support, uniformly in the size of the system. Its
moments grow at most geometrically, they determine the measure
uniquely~\cite{Akhiezer1965}, and the quadratures of \eqref{eq:In} converge weakly
to it~\cite{Gautschi2004}; since $e^{-\lambda t}$ is continuous and bounded on the
support, $I_n(\K,t)$ tends to the decay at every time of the jump regime.
Convergence requires no assumption about the interaction beyond its range.

\section{The measurement}
\label{sec:measured}

An experiment reports the amplitude $B(\K)$ and the rate $\alpha(\K)$ of the
single exponential \eqref{eq:diffusive} fitted in the diffusive regime, and
diffraction gives $F(\K)$. This section identifies these numbers, adds the area beneath the measured decay, and joins both to the theory.

\emph{The fitted rate.} Since $I/F$ is the superposition \eqref{eq:modes} of
exponentials of non-negative weight, $\ln I$ is convex in the time, and the
instantaneous rate
\begin{equation}
-\,\frac{\partial \ln I(\K,t)}{\partial t}
\;=\; \frac{\avg{\lambda\,e^{-\lambda t}}_{\mu}}{\avg{e^{-\lambda t}}_{\mu}}
\label{eq:instrate}
\end{equation}
is \eqref{eq:Omega} extended to every time of the jump regime: a mean of the
rates with the weights $a_k e^{-\lambda_k t}$, equal to $\Omega(\K)$ at $t=0$ and
falling towards $\lambda_{\min}(\K)$ as the memory acts. A fit over a window
$[t_1,t_2]$ returns the mean of this rate,
\begin{equation}
\alpha(\K) \;=\; \frac{\ln I(\K,t_1)-\ln I(\K,t_2)}{t_2-t_1}
\;=\; \frac{1}{t_2-t_1}\int_{t_1}^{t_2}\!
\Bigl[-\frac{\partial \ln I(\K,t)}{\partial t}\Bigr]\,dt ,
\label{eq:alphamean}
\end{equation}
and since the window lies in the diffusive regime, where the instantaneous rate
has settled,
\begin{equation}
\alpha(\K) \;=\; \lambda_{\min}(\K) \;\le\; \Omega(\K) ,
\label{eq:alphalmin}
\end{equation}
when $\lambda_{\min}$ is isolated, and close to the lower edge of the continuum
otherwise. The amplitude $B(\K)$ lies below $F(\K)$, since $\ln I$ is convex, and
is $F(\K)$ times the weight of the slowest mode when this mode is isolated. Neither number carries the part of the decay which precedes the window.
At long wavelength $\lambda_{\min}$ and $\Omega$ nearly coincide, and the
established practice of taking the collective diffusion coefficient from
$\alpha$~\cite{Jardine2009} is exact; elsewhere $\alpha$ describes only the tail
of the decay.

\emph{The correlation time.} The area beneath the normalized decay takes in the whole decay; it is the correlation time,
\begin{equation}
\tau_c(\K) \;=\; \int_0^{\infty}\!\frac{I(\K,t)}{F(\K)}\,dt
\;=\; \frac{\hat I(\K,0)}{F(\K)}
\;=\; \avg{\lambda^{-1}}_{\mu} ,
\label{eq:tauc}
\end{equation}
the second equality being \eqref{eq:spectral} at $s=0$ and the third
\eqref{eq:modes} integrated term by term. It is the mean inverse relaxation rate, independent of any form of the decay; its reciprocal, the harmonic mean of the rates, is not the rate of any stage of the decay. In practice the area is
integrated over the measured points, not fitted, from $t=0$ to the last measured
time $t_{\max}$, and the tail beyond $t_{\max}$ is added from the fit as
$(B/F)\,e^{-\alpha t_{\max}}/\alpha$. At the other end, the ballistic regime and the Brownian equilibration
inside the well last less than a picosecond. If the signal shows no
fast initial drop, the interval from $t=0$, where the signal is $F(\K)$, to the first
measured time $t_1$ is covered by a trapezoid, with an error no larger than
$t_1/\tau_c$. If a fast drop, from the ballistic regime and the equilibration inside the well, is resolved, the slower decay after it is
extrapolated back to $t=0$ and integrated, and its amplitude there replaces
$F(\K)$ in \eqref{eq:tauc} and in every quantity built from the area.

\emph{The ratio $R$.} The local rate and the correlation time are the arithmetic
and the harmonic mean of the same rates, whence
\begin{equation}
R(\K) \;\equiv\; \Omega(\K)\,\tau_c(\K)
\;=\; \avg{\lambda}_{\mu}\avg{\lambda^{-1}}_{\mu} \;\ge\; 1 ,
\label{eq:R}
\end{equation}
with equality only for a single rate; with \eqref{eq:alphalmin}, the rates stand
in the order
$\alpha=\lambda_{\min}\le\tau_c^{-1}\le\Omega$. Setting $s=0$ in
\eqref{eq:laplace} gives the physical meaning of $R$,
\begin{equation}
\frac{1}{\tau_c(\K)} \;=\; \Omega(\K)-\hat\Mc(\K,0) ,
\qquad
R(\K) \;=\; \Bigl[\,1-\frac{\hat\Mc(\K,0)}{\Omega(\K)}\,\Bigr]^{-1} :
\label{eq:Rmemory}
\end{equation}
the memory, integrated over the whole jump regime, lowers the rate from $\Omega$ to
$\tau_c^{-1}$, and $R$ is the factor by which it lengthens the correlation time. Consequently a decay without memory underestimates the correlation time by the
factor $R$, and a hop rate taken from the area as if there were no memory is
$\Gamma_{\rm eff}/R$.
Each order $n$ of the hierarchy gives its own value $R^{(n)}$ from the $n$-point
spectrum of \eqref{eq:In}; the first places the whole weight at $\Omega$, and Gauss
quadrature underestimates $\avg{\lambda^{-1}}_{\mu}$ at every order, so that
\begin{equation}
1 \;=\; R^{(1)} \;\le\; R^{(2)} \;\le\; \cdots \;\le\; R .
\label{eq:Rladder}
\end{equation}
The order at which $R^{(n)}$ reaches the measured $R$ fixes how many moments the
hierarchy needs. $R$ and $\Delta$ are independent, $\Delta$ weighing the fast rates
and $R$ the slow ones: a weight $10^{-3}$ at a rate $10^{3}$ times larger than the
rest gives $\Delta=249.5$ and $R=1.997$, and at a rate $10^{6}$ times smaller
$\Delta=10^{-3}$ and $R=1000$.

\begin{figure}[!ht]
\centering
\includegraphics[width=0.99\textwidth]{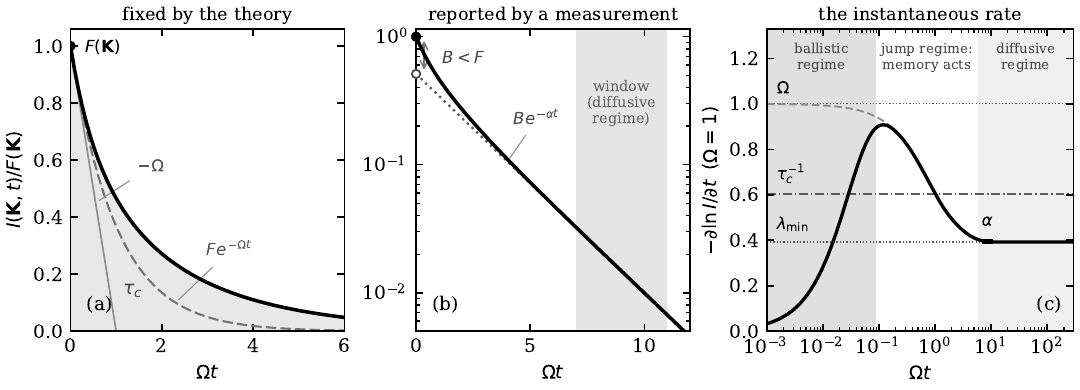}
\caption{The quantities of both kinds on one decay, the exact intermediate
scattering function of an arbitrary admissible spectrum normalized to $\Omega=1$;
no model of the adlayer enters. The time is measured from the instant of
scattering. (a) The theory: the amplitude $F$, the initial slope $-\Omega$, the
first order $Fe^{-\Omega t}$, and the correlation time $\tau_c$ as the shaded area.
(b) The measurement, on a logarithmic scale: the single exponential fitted in a
window in the diffusive regime, with $B<F$ and $\alpha=\lambda_{\min}$. (c) The
instantaneous rate \eqref{eq:instrate}, with a schematic ballistic regime: it
vanishes at the origin, approaches $\Omega$ after the ballistic regime and falls,
as the memory acts, to $\lambda_{\min}$, where $\alpha$ (thick segment) is
measured; dashed, the decay of the jump regime continued to $t=0$.}
\label{fig:anatomy}
\end{figure}

Figure~\ref{fig:anatomy} shows these quantities on one decay.

\emph{Use of the data.} Comparing $B$ with $F$ settles whether the decay is a
single exponential: if they agree within errors, $\alpha=\tau_c^{-1}$ and the
established analysis is complete; if not, neither $1/\alpha$ nor $B/\alpha$ is the
correlation time. The area assumes that the run reaches the diffusive regime, $\lambda_{\min}$ exceeding the reciprocal of the longest time measured. A slower mode would be absorbed into the elastic baseline $A(\K)$ of \eqref{eq:combsplit} without any sign in the data: a weight $10^{-2}$ at a rate one thousand times smaller makes the true $\tau_c$ eleven times the visible area, with $B=0.99\,F$.
A longer run at the same wavevector tests it: the area must not grow.

The structure factor, the correlation time and the single-jump CF $f_{\K}$ enter the theory in one combination: the sum rule \eqref{eq:sumrule}, multiplied by $\tau_c$ and divided by $F$, gives
\begin{equation}
R(\K) \;=\; \Gamma_{\rm eff}\,
\frac{\bigl[1-f_{\K}\bigr]\,\tau_c(\K)}{F(\K)} ,
\label{eq:Rmeas}
\end{equation}
measured quantities and the lattice on the right, one scalar $\Gamma_{\rm eff}$,
and on the left a quantity which cannot fall below unity. The procedure has four
steps and no adjustable parameter.
\begin{enumerate}[label=(\roman*),leftmargin=2.6em]
\item \emph{Integrate.} At every wavevector form
$F(\K)/\{[1-f_{\K}]\,\tau_c(\K)\}$, with no fit and no window.
\item \emph{Fix the scalar.} For dynamics of gradient type, in the sense of
Sec.~\ref{sec:intro}, $R(0)=1$, and \eqref{eq:Rmeas} gives
\begin{equation}
\Gamma_{\rm eff} \;=\; \lim_{K\to0}
\frac{F(\K)}{\bigl[1-f_{\K}\bigr]\,\tau_c(\K)} .
\label{eq:protocol}
\end{equation}
\item \emph{Test.} No value of the quotient of step (i) may exceed this limit,
since $R$ would there fall below unity.
\item \emph{Obtain the rate and the memory.} Equation \eqref{eq:Rmeas} gives
$R(\K)$ across the zone and the local rate $\Omega=R/\tau_c$, which no fit reaches;
\eqref{eq:Rladder} says how many moments the hierarchy needs, and the static
averages of Sec.~\ref{sec:stieltjes} supply them.
\end{enumerate}
The only new operation the procedure asks of an experiment is the integration of
the decay, from data a spin-echo experiment already takes.

\subsection*{The hydrodynamic limit}

For dynamics of gradient type the adlayer behaves at long wavelength as a medium
with one transport coefficient~\cite{Spohn1991,KipnisLandim1999}, and
\begin{equation}
\lim_{K\to0}R(\K) \;=\; 1 ,
\qquad\text{so that}\qquad
D_c \;=\; \frac{\Gamma_{\rm eff}\,\avg{\ell^{2}}}{2\,F(0)}
\quad\text{exactly, at every coupling.}
\label{eq:R0}
\end{equation}
The first order \eqref{eq:level1} then gives $D_c$ exactly at long wavelength,
although the memory does not vanish at finite $\K$; the coefficient of the local
rate, \eqref{eq:Dc1}, is also that of the diffusive regime, \eqref{eq:kappa2}; and
\eqref{eq:protocol} gives $\Gamma_{\rm eff}$ itself. The memory and \eqref{eq:R0}
are compatible because almost the whole weight of the spectrum sits at rates of
order $K^{2}$, which set $\tau_c$, while a vanishing remainder stays at rates of
the order of the hop rate and keeps $\Delta(0)$ finite. Where the dynamics is not
of gradient type, $R(0)$ exceeds unity, the coefficient of the diffusive regime is
$D_c/R(0)$, \eqref{eq:protocol} gives $\Gamma_{\rm eff}/R(0)$, and $R(0)$ is the reciprocal of the collective correlation factor
of the lattice-gas literature~\cite{AlaNissila2002}.

\section{Concluding remarks}
\label{sec:concl}

The local rate and the memory function have been reduced to static averages of the
adlayer, for a class of lattice gases with an unspecified interaction: the first
through the sum rule, the second, order by order, through the continued fraction
its moments generate.

The experiment records the signal from the instant of scattering. The theory
describes it from the end of the ballistic regime, through the jump regime in which
the memory acts, to the diffusive regime of dynamical equilibrium; one memory equation covers both, and the ballistic regime and the equilibration inside a well lie outside it. On this common
time axis the local rate is the instantaneous rate at the beginning of the jump
regime, and no fit reaches it. The memory then lowers the rate: the slowest rate
$\lambda_{\min}$ is the constant rate it leaves in the diffusive regime, and the
inverse correlation time is the local rate lowered by the memory integrated over
the whole jump regime, \eqref{eq:Rmemory}. The rate a fit reports, from a window
in the diffusive regime, is $\lambda_{\min}$, below the inverse correlation time
and the local rate. The correlation time is the area beneath the decay, and its
reciprocal is an average over the whole decay, not the rate of any stage. At long
wavelength one mode carries almost the whole weight, the three rates nearly
coincide, and the established analysis is exact.

Two qualifications apply. Reversibility does not produce the sum rule. The real part \eqref{eq:dirichlet} holds by stationarity alone, and detailed balance only removes an antisymmetric contribution which no combination of the exchange frequencies reproduces; an adlayer held out of equilibrium keeps the first and loses the second. And the separation of the rate into a structureless numerator and a structured denominator, the lattice form of the narrowing of de Gennes~\cite{deGennes1959}, requires jump vectors equivalent under the symmetry of the lattice. Otherwise the weights \eqref{eq:weights} are ratios of exchange frequencies which the interaction reweights, and part of the momentum dependence is no longer geometrical. At long wavelength the same separation becomes the
relation of Darken, with the rate the sum rule supplies in place of the mobility of
a labelled adparticle. The relation fixes the coefficient of the local rate; for
dynamics of gradient type it is also the coefficient of the diffusive regime, and
otherwise the two differ by $R(0)$.

Detailed balance also makes the decay of the jump regime a superposition of
relaxation modes of non-negative weight, and with it gives the hierarchy
\eqref{eq:cf}, each order fixed by static averages and lying below the decay; the
closed form \eqref{eq:level1} is its first member. The gain from higher orders is
governed by $\Delta$, which weighs the fast rates, and by $R$, which weighs the
slow ones; neither bounds the other.

A measurement reaches $R$, from data already taken. With the structure factor and
the single-jump geometry, the area beneath the decay forms the quotient
$F(\K)/[(1-f_{\K})\tau_c(\K)]$, which the theory identifies with
$\Gamma_{\rm eff}/R(\K)$. Since $R$ never falls below unity, this quotient can
nowhere exceed its long-wavelength value: a test with no adjustable parameter. For
dynamics of gradient type this value is the effective hop rate, obtained with no
model of the adlayer, no fit, and no assumption that the decay is exponential. The
one assumption, that no part of the decay is hidden in the elastic baseline, is
tested by a longer run, which also checks that the measurement reaches the
diffusive regime. The local rate and $R$ are then available across the zone, and
the departure of $R$ from unity says, through \eqref{eq:Rladder}, how many moments
the hierarchy needs.

Four matters are left open. The effective rate is defined but not evaluated; its
evaluation for a given adlayer is the point at which a model must be chosen. The
hierarchy converges for the whole class, but how fast depends on the spectrum, and
so on the interaction. The separation of the spectrum into a hydrodynamic part and
a fast remainder has been described by its consequences alone; a quantitative
account of how the fast weight vanishes with the wavevector would require the
separation to be defined by the spectrum itself. The effective hop rate requires
dynamics of gradient type, which rates fixed by the departure site satisfy in one
dimension but not in general on a two-dimensional lattice; there the procedure
gives $\Gamma_{\rm eff}/R(0)$, and $R(0)$ must be obtained otherwise.

\section*{Acknowledgements}

This work was supported by the Ministry of Science, Innovation and Universities
(Spain) under Grant No. PID2023-149406NB-I00.

\section*{Declaration of competing interests}

The author declares that he has no known competing financial interests or
personal relationships that could have appeared to influence the work reported
in this article.

\section*{Data availability}

No datasets were generated or analysed in this study. The curve of
Fig.~\ref{fig:anatomy} is computed from the relaxation spectrum its caption
describes.

\appendix

\section{The first orders}
\label{app:levels}

The coefficients of \eqref{eq:cf} are generated by the moments
\eqref{eq:moments} through the three-term recurrence of the polynomials
orthogonal with respect to $\mu_{\K}$, and are written most compactly with the central moments of the relaxation spectrum,
\begin{equation}
\bar m_n(\K) \;=\; \bigl\langle\,\bigl[\lambda-\Omega(\K)\bigr]^{n}\,
\bigr\rangle_{\mu} ,
\label{eq:central}
\end{equation}
of which $\bar m_2$ is the variance of the rates, $\bar m_3$ their skewness
before normalization, and $\bar m_1=0$ by the definition of $\Omega$. The first
coefficients are
\begin{equation}
\Omega_0 \;=\; \Omega , \qquad
b_1 \;=\; \bar m_2 , \qquad
\Omega_1 \;=\; \Omega+\frac{\bar m_3}{\bar m_2} , \qquad
b_2 \;=\; \frac{\bar m_4}{\bar m_2}-\bar m_2-\frac{\bar m_3^{2}}{\bar m_2^{2}} ,
\label{eq:coeffs1}
\end{equation}
\begin{equation}
\Omega_2 \;=\; \Omega \;+\;
\frac{\bar m_2^{2}\bar m_5-2\bar m_2\bar m_3\bar m_4+\bar m_3^{3}}
{\bar m_2^{3}\,b_2} .
\label{eq:coeffs2}
\end{equation}
Order $n$ requires the moments up to $m_{2n-1}$. The
squared relative width of \eqref{eq:Delta} is $\Delta=b_1/\Omega^{2}$, and
$b_2\ge0$ is the non-negativity of the determinant
$\bar m_2\bar m_4-\bar m_2^{3}-\bar m_3^{2}$, one of the conditions a sequence
of numbers must satisfy to be the moments of a non-negative
measure~\cite{Akhiezer1965}.

Evaluating \eqref{eq:cf} at $s=0$ gives the correlation time at each order, and
with it the ratio \eqref{eq:R},
\begin{equation}
R^{(1)} \;=\; 1 , \qquad
R^{(2)} \;=\; \Bigl[\,1-\frac{b_1}{\Omega_0\,\Omega_1}\,\Bigr]^{-1} ,
\qquad
R^{(3)} \;=\; \Bigl[\,1-\frac{b_1}
{\Omega_0\bigl(\Omega_1-b_2/\Omega_2\bigr)}\,\Bigr]^{-1} ,
\label{eq:Rlevels}
\end{equation}
each obtained from the preceding one by continuing the fraction one step. $R^{(1)}=1$ because the first order places the whole spectrum at $\Omega$, where the arithmetic and harmonic means coincide; every further order subtracts a non-negative quantity from the denominator and increases the ratio, which is \eqref{eq:Rladder}.

The corresponding closed forms for the memory function are obtained from
\eqref{eq:memcf}. At the second order it is the single exponential
\eqref{eq:memlevel2}, whose time integral is $b_1/\Omega_1$; at the third order
it is
\begin{equation}
\Mc(\K,\tau) \;=\; b_1\,\frac{(\nu_+ -\Omega_2)\,e^{-\nu_+\tau}
-(\nu_- -\Omega_2)\,e^{-\nu_-\tau}}{\nu_+-\nu_-} ,
\qquad
\nu_{\pm} \;=\; \frac{\Omega_1+\Omega_2}{2}
\pm\sqrt{\Bigl(\frac{\Omega_1-\Omega_2}{2}\Bigr)^{2}+b_2} ,
\label{eq:memlevel3}
\end{equation}
which starts at the exact initial value $b_1$ and is non-negative throughout, $b_2\ge0$ placing $\nu_-$ below $\Omega_2$ and $\nu_+$ above it. Each order keeps the two properties of the exact memory, a fixed initial value and a definite sign, and differs from it only in the shape of its decay.

\end{document}